\documentclass{pasj00}

\newcommand{\average}[1]{\ensuremath{\langle#1\rangle} }

\usepackage{color}
\usepackage{ulem}
\usepackage{times,bm}
\usepackage{multicol}

\Received{yyyy/mm/dd}
\Accepted{yyyy/mm/dd}
\Published{$\langle$publication date$\rangle$}
\SetRunningHead{S.\ Takeuchi, K.\ Ohsuga, and S. Mineshige}{
RHD Instability in 
Plane-Parallel, Super-Eddington Atmosphere
}

\begin{document}

\title{Radiation Hydrodynamic Instability in
Plane-Parallel, Super-Eddington Atmosphere:
A Mechanism for Clump Formation
}
\author{Shun \textsc{Takeuchi},\altaffilmark{1}\thanks{Present address:  Fujitsu Limited, 1-9-3 Nakase, Mihama-ku, Chiba
261-8588.}
       Ken \textsc{Ohsuga},\altaffilmark{2, 3}
        and
        Shin \textsc{Mineshige}\altaffilmark{1}
        }
\altaffiltext{1}{Department of Astronomy, Graduate School of Science, Kyoto University, Kitashirakawa-Oiwake-cho, Sakyo-ku, Kyoto 606-8502}
\email{shun@kusastro.kyoto-u.ac.jp}
\altaffiltext{2}{National Astronomical Observatory of Japan, 2-21-1 Osawa, Mitaka, Tokyo 181-8588}
\altaffiltext{3}{School of Physical Sciences, Graduate University of Advanced Study (SOKENDAI), Shonan Village, Hayama, Kanagawa 240-0193}

\maketitle

\begin{abstract}
In order to understand the physical processes underlying clump formation in outflow from
supercritical accretion flow,
we performed two-dimensional radiation hydrodynamic (RHD) simulations. 
We focus our discussion on the nature of RHD instability in 
marginally optically thick,
plane-parallel, super-Eddington atmosphere.
Initially we set two-layered atmosphere with a density contrast of 100
exposed to strong, upward 
continuum-radiation force;
the lower layer is denser than the upper one, 
condition for an RHD instability.
We assume non-zero but negligible gravitational force, compared with the radiation force.
We find that short wavelength perturbations first grow, 
followed by growth of longer wavelength patterns, 
which lead to the formation of clumpy structure.
The typical size of clumps (clouds)
corresponds to about one optical depth. 
An anti-correlation between the radiation pressure and the gas pressure is confirmed:
this anti-correlation provides a damping mechanism
of longer wavelength perturbations 
than the typical clump size. Matter and radiation energy densities are correlated.
These features are exactly 
what we found in the radiation-magnetohydrodynamic 
(radiation-MHD)
simulations of supercritical outflow.
\end{abstract}

\section{Introduction}
Radiation fields are known to play a number of important roles in astrophysics.
They do not only carry information from space to observers
but can also contribute to the energetics, and even to the dynamics, of astrophysical phenomena.
Matter emits, absorbs, and scatters radiation, while radiation gives (or removes) 
energy and momentum to (from) matter.
Such radiation-matter interactions are one of the most important issues in astrophysics,
since unique types of instabilities and associated active phenomena could take place.

The most important key parameter when we consider dynamics of radiating objects is the
Eddington parameter $\Gamma$,
ratio of the outward acceleration by radiation to the inward acceleration by gravity.
Luminous objects shining above the Eddington limit, $\Gamma > 1$, are of particular interest.
Their most notable feature is the emergence of radiation-driven outflow;
it can cause dynamical feedback to their environments,
and sometimes has a large impact on the evolution of the surrounding media
(\cite{Wan+06}; \cite{Kru+09}; \cite{Kro+09}).
Such unique features of super-Eddington object have been extensively discussed 
in various astrophysical contexts,
including luminous blue variables (LBVs), Wolf-Rayet stars, classical novae, 
supernovae, microquasars, active galactic nuclei (AGNs), and so on
(\cite{DavHum97}; \cite{Min+00}; \cite{NugLam00}; \cite{Sha00}; \cite{Rev+02}; \cite{Smi+09}).

Supercritical (or super-Eddington) accretion flow
is a model for ultraluminous X-ray sources (ULXs),
luminous microquasar and AGN with $\Gamma > 1$
(\cite{ShaSun73}; \cite{Egg+88}; \cite{Wan+99}
\cite{OkuFuj00}; \cite{Wat+01}).
Multi-dimensional radiation-hydrodynamic (RHD) 
and radiation-magnetohydrodynamic (radiation-MHD) simulations
of supercritical sources have confirmed
that steady, supercritical accretion onto black holes is feasible,
as long as accretion occurs through a disk (\cite{Ohs+05}; \cite{Ohs+09}).
Continuous radiation from supercritical accretion flows
drives outflow, by which significant amount of matter is blown away
(\cite{Fuk04}; \cite{Tak+09}; \cite{Tak+10}; \cite{KruTho12}; \cite{KruTho13}).
 
Here, we pay attention to the outflow structure itself,
instead of its environmental effects.
\citet{Tak+13} (hearafter Paper~I) reported the emergence of clumpy outflow
from supercritical accretion flow onto a black hole by means of global two-dimensional
radiation-MHD simulations.
The typical size of the clumps (clouds)
is $\sim 10 r_{\rm S}$
(with $r_{\rm S}$ being the Schwarzschild radius),
which corresponds to about one optical depth.
The presence of clumpy features has been independently indicated to 
account for significant time variabilities in the observations of 
luminous accretion flow (\cite{Fab04}; \cite{Mid+11}; \cite{Tom+12}).
Clumpy outflow was also considered in relation to the AGN unified model
to explain the origin of the broad-line region (BLR) clouds.
\citet{Eli12}, for example, proposed that the BLR clouds originate from 
clumpy outflow gas flowing around luminous AGNs.
His model can nicely explain the observed BLR disappearance at low luminosity 
(\cite{Nic00}; \cite{EliHo09}).

Although the clumpy outflow was nicely demonstrated by the simulations,
we did not specify responsible physical mechanisms of clump formation in Paper I. 
Radiation processes should be somehow involved,
since the typical clump size is regulated by the optical depth,
whereas magnetic processes cannot be essential
since similar clumpy structure is found in non-magnetic RHD simulation data.
In paper I, therefore, we tentatively concluded that the Rayleigh-Taylor (RT) instability
by strong continuum-radiation force is 
the most plausible cause of clump formation.

While radiation-driven RT instability is a classical issue (\cite{MatBlu+77}; \cite{Kro77}),
this subject is again attracting astrophysicists quite recently.
\citet{JacKru11} performed linear stability analysis of the RT instability in radiating fluids.
Their main conclusions can be summarized as follows:
In the optically thin limit,
radiation field can be expressed as the external field
and acts as part of an effective gravitational field.
As a result, the dispersion relation of the instability is derived by
the generalized formula with RHD
and the instability criterion can be explained in terms not of the matter density 
but of the momentum of two media.
In the optically thick limit, on the other hand,
the dispersion relation eventuates that of the pure hydrodynamic RT instability
because of strong coupling between matter and radiation.
Rather extensive, large-scale numerical studies have started only recently
after the rapid developments of high-speed supercomputers
and the improvements in the formulation of RHD and in the numerical technique.
By means of two-dimensional RHD simulations \citet{Jia+13} investigated the flow properties
of two optically thick, uniform layers with different densities under hydrostatic equilibrium
between the gravity force and the radiation-pressure force.
They reported that the RT instability occurs even in an optically thick medium,
but the growth of small scale perturbations are suppressed by the radiation filed.
We should note, however, that these previous studies of the radiation RT instability 
were made under the assumption of sub-Eddington atmosphere.

Unlike the cases with sub-Eddington outflow with $\Gamma < 1$, 
the direction of the net acceleration is opposite to that of the
gravitational force in super-Eddington outflow with $\Gamma > 1$.
A system is thus dynamical unstable, when lighter fluid lies above heavy fluid ({\cite{Cha61}),
situation which more easily realizes, especially when outflow goes out to wider directions.
In this paper, we explore RHD instability
as a possible cause of clump formation in super-Eddington atmosphere.
For this purpose, we postulate a rather simple case; that is, we set a
two-layered, plane-parallel, marginally optically thick atmosphere under constant gravity,
although in reality gravitational force may not be uniform and advective and/or convective gas motions
can be associated, as we saw in Paper I.
Our goal is to give satisfactory explanation concerning 
the origin of clumpy outflow and the nature of the associated instability.

The plan of this paper is as follows. 
In the next section, we overview the basic equations and 
the simulation model of the RHD instability.
We present the result and the discussion in section 3 and 4.
The final section is devoted to concluding remarks.
It should be noted that we avoid to use the terminology of the radiation RT instability
in the present study, since 
gravity is not essential for the instability,
though physical processes look similar.

\section{Our Model and Numerical Procedures}
\subsection{Overview}
In the present study we postulate a rather simplified situation,
not to lose but to capture the essence of the RHD instabilities
in super-Eddington atmosphere.
We simulate time developments of small density perturbations 
(caused by initially added, small velocity perturbations)
embedded in a two-layered, marginally optically thick, super-Eddington atmosphere
exposed to strong, upward continuum-radiation force.
The lower layer is denser than the upper one, 
and set the density contrast to be 100.
We assume a constant gravity, but its magnitude is negligible, compared with the radiation force.
We further assume no magnetic fields, 
since magnetic fields are not essential for the clump formation (see Paper~I).

We base the numerical code used in Paper~I to compare with the clumpy outflow.
The code is a two-dimensional radiation-MHD solver of black-hole accretion flow
by the modified Lax-Wendroff scheme,
which is used in cylindrical coordinates $(r, \theta, z)$ 
(see also \cite{OhsMin11} for details).
However, we drop the curvature terms by setting $1/r \rightarrow 0$.
In short, we examine the developments of
the non-magnetic RHD instabilities in plane-parallel super-Eddington
atmosphere by using the Cartesian coordinates $(x, y, z)$.

\subsection{Basic Equations}
The properties of radiating fluids are described by the combination of the
hydrodynamic equation and the equations for radiation.
In this paper, we consider that fluids are non-dissipative and compressive;
that is, the behavior of the fluid obeys the Euler's equation.
The basic equations that contain the terms up to the order of $(v/c)^1$ are 
the continuity equation,
the momentum equation of matter,
the internal energy equation of matter,
and the radiation energy equation:
\begin{equation}
\frac{\partial \rho}{\partial t}
+ {\bm \nabla} \cdot (\rho {\bm v}) = 0,
\label{mass_con}
\end{equation}
\begin{eqnarray}
\frac{\partial (\rho {\bm v})}{\partial t} + {\bm \nabla} \cdot
\left( \rho {\bm v}{\bm v} \right) = - {\bm \nabla}  p_{\rm gas}
 + \frac{\kappa_{\rm es} \rho}{c} {\bm F}_0 + \rho \textsl{\bf g},
\label{mom}
\end{eqnarray}
\begin{eqnarray}
\frac{\partial e }{\partial t}
+ {\bm \nabla}\cdot(e {\bm v})  = - p_{\rm gas} \left( {\bm \nabla}\cdot{\bm v} \right),
\label{gase}
\end{eqnarray}
\begin{eqnarray}
\frac{\partial E_0}{\partial t} + {\bm \nabla}\cdot(E_0 {\bm v}) 
 = - {\bm \nabla} \cdot {\bm F_0} - {{\bf P}_0}:{\bm \nabla} {\bm v},
\label{rade}
\end{eqnarray}
(\cite{MihMih84}).
Here, $\rho$ is the matter density,
$\bm{v}$ is the flow velocity,
$c$ is the speed of light,
$e$ [$={p_{\rm gas}}/({\gamma -1})$] is the internal energy density of matter,
$p_{\rm gas}$ is the gas pressure,
$\textsl{\bf g}$ is the gravitational acceleration,
$E_0$ is the radiation energy density,
${\bm F}_0$ is the radiative flux,
${ {\bf P}}_0$ is the radiation-pressure tensor,
$\gamma$ is the ratio of the specific heats and we set $\gamma=5/3$,
$\kappa_{\rm es}$ ($=\sigma_{\rm T}/m_{\rm p}$) is the electron scattering opacity, 
$\sigma_{\rm T}$ is the Thomson scattering cross-section,
and $m_{\rm p}$ is the proton mass, respectively.
The subscript 0 for radiation filed means the value measured
in the co-moving (fluid) frame. For the sake of simplicity,
we adopt the gray (frequency-integrated) approximation for the
radiation terms.

The set of equations (\ref{mass_con})-(\ref{rade}) is closed 
by an ideal gas equation of state,
\begin{equation}
p_{\rm gas} = \frac{k_{\rm B}}{\mu m_{\rm p}} \rho T_{\rm gas},
\end{equation}
and 
by adopting the flux-limited diffusion (FLD) approximation
to evaluate $\bm F_0$ and ${\bf P}_0$ (\cite{LevPom81}),
\begin{eqnarray}
 \bm{F}_0 = -\frac{c\lambda}{\kappa_{\rm es} \rho} \nabla E_0,
\label{gase}
\end{eqnarray}
\begin{eqnarray}
 {\bf P}_0 = {\bf f} E_0,
\label{closuer}
\end{eqnarray}
where
\begin{eqnarray}
 \lambda = \frac{2+{\cal R}}{6+3{\cal R}+{\cal R}^2},
\end{eqnarray}
\begin{eqnarray}
 {\bf f} = \frac{1}{2}(1-f){\bf I} + \frac{1}{2} (3f-1)\bm{n}\bm{n},
\end{eqnarray}
\begin{eqnarray}
 f =  \lambda + \lambda^2 {\cal R}^2,
\end{eqnarray}
and
$k_{\rm B}$ is the Boltzmann constant,
$\mu$ is the mean molecular weight, 
$T_{\rm gas}$ is the temperature of the gas,
$\lambda$ is the flux limiter, 
$\bf f$ is the Eddington tensor, 
$f$ is the Eddington factor,
${\cal R}$~[$= |\nabla E_0|/(\kappa_{\rm es}\rho E_0)$] is the dimensionless quantity, and
$\bm{n}$~($= \nabla E_0/|\nabla E_0|$) is 
the unit vector in the direction of the radiation energy density gradient,
respectively.
We assume that the fluid consists of the fully ionized hydrogen
($\mu = 0.5$),
although \citet{JacKru11} consider dependence on the chemical composition.

In this simulation, we assume constant gravity field in the vertical direction,
\begin{eqnarray}
\textsl{\bf g} = - \mathrm{g}_0 \bm{e}_z,
\label{gra}
\end{eqnarray}
where $\mathrm{g}_0$ ($\equiv 10 \kappa_{\rm es} \rho_0 c_{\rm i,b}^2/\gamma$)
is the gravitational acceleration which is assumed to be constant,
$\rho_0$ a constant matter density,
and $c_{\rm i,b}$ ($\equiv 10^{-3} c$) the initial isothermal sound speed
at the bottom of the calculation box.

\subsection{Initial Conditions}
Initially, the lower domain ($z < 0$) is filled with optically thick, 
heavy fluid of matter density, $\rho_- = \rho_0$,
while the upper domain ($z > 0$) is filled with optically thin, 
light fluid of matter density, $\rho_+ = 10^{-2} \rho_0$.
The interface of the density jump at $z = 0$
is connected by the hyperbolic function.
The Atwood number is expressed by
$A = (\rho_- - \rho_+)/(\rho_- + \rho_+) = 0.99/1.01$.
Note that the configuration of the two uniform fluids in this simulation
is counter to traditional RT instability,
in which the heavy fluid lies above the light one.

Generally, both dynamical and thermal equilibrium as the initial condition
should be satisfied to study RT instability.
However, as \citet{Jia+13} pointed out,
the dynamical equilibrium
cannot be achieved
when the thermal equilibrium is realized.
The radiation energy density will jump with the matter density jump at the interface
due to the strong coupling between radiation and matter,
$E_0 \propto T_{\rm gas}^4 \propto (p_{\rm gas}/\rho)^4$.
Therefore, we neglect the emission and absorption in this paper.
Since the thermal timescale is much longer than 
the growth time of the instabilities of interest,
we may assume that the initial state is out of thermal equilibrium.

We consider moving atmosphere at a constant speed, for simplicity, 
and describe the basic equations in the co-moving frame of the moving atmosphere.
In this frame the dynamical equilibrium is satisfied at $t=0$;
\begin{equation}
\frac{dp_{\rm gas}}{d x} = F^x_0 = 0,
\label{dyn_eqx}
\end{equation}
\begin{equation}
- \frac{dp_{\rm gas}}{d z} + \frac{\kappa_{\rm es} \rho}{c} F^z_0 
- \rho \mathrm{g}_0 = 0.
\label{dyn_eqz}
\end{equation}
From equation (\ref{dyn_eqz})
the initial vertical profile of the gas pressure is calculated by 
\begin{equation}
p_{\rm gas} = p_{\rm gas,b}
+ \int \left( \Gamma  - 1 \right)  f_{\rm gra}^z dz,
\end{equation}
where 
\begin{equation}
\Gamma 
\equiv \frac{f_{\rm rad}^z}{f_{\rm gra}^z} 
= \frac{\kappa_{\rm es}F^z_0}{c \mathrm{g}_0}.
\label{eddpara}
\end{equation}
and $p_{\rm gas,b}$ ($= \rho_0 c_{\rm i,b}^2$) is the gas pressure at the bottom,
$\Gamma$ is again the Eddington parameter, and
$f_{\rm rad}^z$ and $f_{\rm gra}^z$ are 
the vertical force of the radiation and the gravity, respectively.
Figure \ref{initial} shows 
the initial vertical profiles of the normalised gas mass density (solid), $\rho/\rho_{\rm b}$,
the normalised gas temperature (dashed), $T_{\rm gas}/T_{\rm gas,b}$, and
the ratio of the radiation energy density and the internal energy density (dotted), $E_0/e$.

\begin{figure}[!t]
\begin{center}
\FigureFile(85mm,55mm){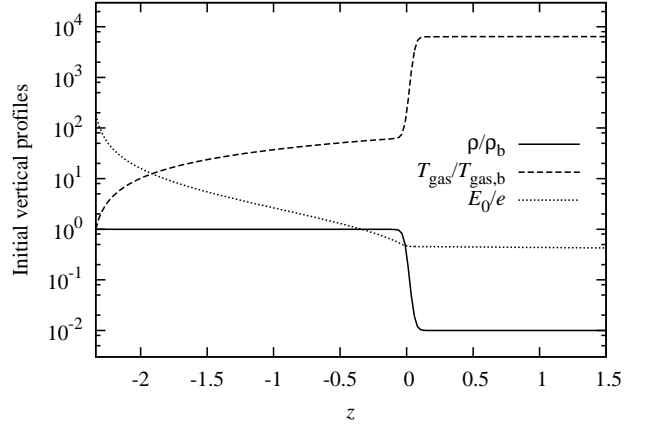}
\caption{
The initial vertical profiles of the normalised gas mass density (solid),
the normalised gas temperature (dashed), and
the ratio of the radiation energy density and the internal energy density (dotted).
}
\label{initial}
\end{center}
\end{figure}

For the setup of RHD instability,
perturbations are added to the vertical velocity component 
throughout the computational domain.
We set the amplitudes of the perturbations to be kept small compared to the sound speed
and to decrease toward the vertical boundaries; thus,
 \begin{eqnarray}
  v_z =
  \left\{ \begin{array}{ll}
    \displaystyle  v_0 \exp{\left( k_z z \right)}
          & \displaystyle (z < 0) \\
     \displaystyle  v_0 \exp{\left( - k_z z \right)}
          & \displaystyle (z \ge 0),
  \end{array} \right.
\label{perturb}
\end{eqnarray}
where $v_0$ $ = 10^{-2} R c_{\rm i,b}$,
$k_z$ ($= 2\pi/z_{\rm l}$) is the wave number, 
$z_{\rm l}$ is the width of the lower domain,
and $R$ is the random number between $-1$ and $1$.
The perturbed velocities are at most $1\%$ 
of the sound speed in the heavy fluid.

\subsection{Grids and Boundary Conditions}
The computational domain is $-6.02 \le x \le 6.02$ and $-2.36 \le z \le 5.18$,
where $x$ and $z$ are normalized by the photon mean free path
in the lower domain,
\begin{equation}
\ell = \frac{1}{\kappa_{\rm es} \rho_0}.
\label{pmfp}
\end{equation}
It is divided into $514 \times 322$ cells ($x \times z$) with a constant grid size.
The optical depths of the lower heavier layer 
in the horizontal ($x$) and vertical ($z$) directions 
are $\tau_x \sim 12$ and $\tau_z \sim 2$, respectively,
while the upper lighter layer is optically thin
because of much smaller density, $\rho_+ = 10^{-2}\rho_0$.

We adopt reflecting boundary at all the boundaries.
In the traditional numerical study of RT instability,
periodic boundary conditions are adopted in the $x$ direction.
By setting a fairly large domain in the $x$ direction,
we can focus on the instability in the inner domain.
Note that the computational boundaries can hardly affect
the structural changes, at least, in the central part of the simulation box, since sound wave can propagate over a distance of 1.95 during the computation time of 1.95
(in the unit of the sound crossing timescale over the photon mean free path,
normalization constant of the spatial coordinates), 
while the horizontal size of the simulation box is $12.04$.
The vertical components of the radiative flux 
are set to be $cE_0$ at the upper boundaries,
and the radial components of the radiative
fluxes are set to be zero at the inner and outer boundaries. 
We have a constant radiation energy density at the lower boundary
which is determined so as to attain $\Gamma = 10$ at $t = 0$ in this paper;
that is, the atmosphere is super-Eddington.

\section{Results}
\subsection{Evolution of the Radiation Hydrodynamic Instability}
\begin{figure}[!ht]
\begin{center}
\FigureFile(85mm,55mm){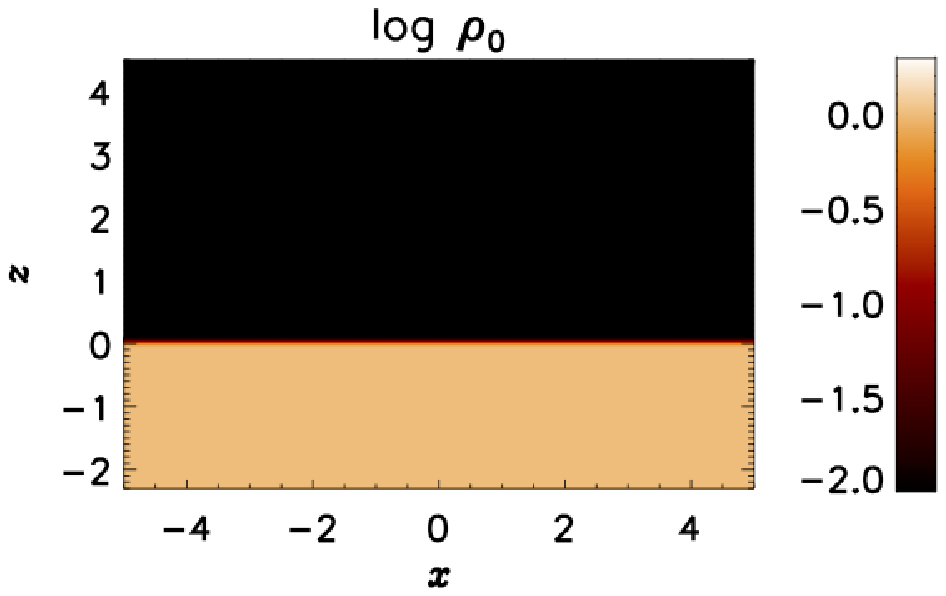}
\FigureFile(85mm,55mm){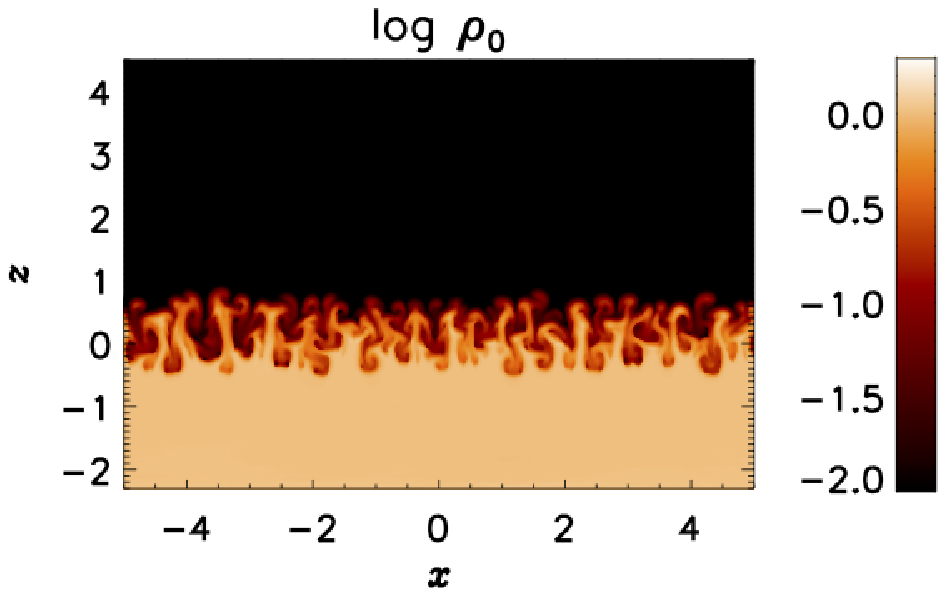}
\FigureFile(85mm,55mm){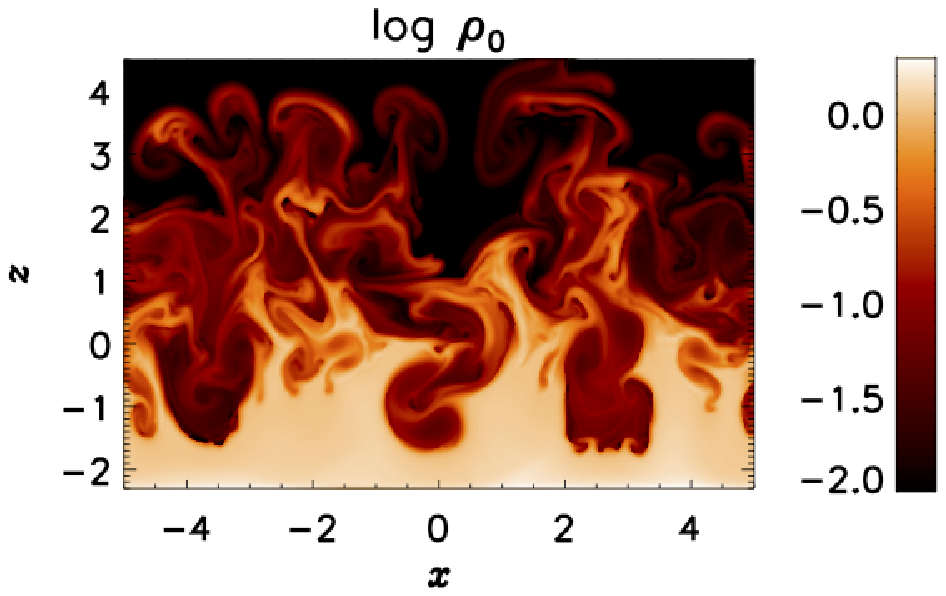}
\caption{
Time evolutionary sequence of an RHD instability 
in plane parallel, super-Eddington atmosphere.
Color contours represent the matter density
at elapsed times of $t = 0$ (top), 
$t = 0.90$ (middle), and
$t = 1.95$ (bottom)
in the unit of sound crossing time over photon mean free path.
(Color online)
}
\label{evolution}
\end{center}
\end{figure}

\begin{figure}[!t]
\begin{center}
\FigureFile(85mm,55mm){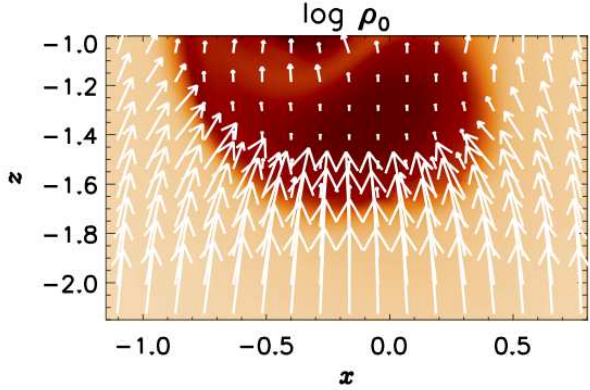}
\caption{
Magnified view showing the interface of an optically thick spike.
Color contours representing matter density are overlaid with the radiation
force vectors per unit volume (arrows) 
at the elapsed time of $t = 1.95$
(corresponding to the bottom panel in Figure \ref{evolution}).
There is a clear tendency that the force vectors mainly direct from denser region 
to less dense region. 
Also, the magnitude of the radiation force is correlated with the matter density.
(Color online)
}
\label{frad}
\end{center}
\end{figure}

Let us show first how an RHD instability grows in super-Eddington atmosphere in
Figure \ref{evolution}, which shows the matter density contours
at the elapsed times of $t = 0$ (top panel), $t = 0.90$ (middle panel), 
and $t = 1.95$ (bottom panel), respectively.
The elapsed times are normalized by the sound crossing timescale 
over photon mean free path.
$t_{\rm sc} \equiv \ell/c_{\rm i,b}$. 
The upward continuum radiation force balances with
the sum of the downward gravitational and gas-pressure forces in the initial state (top),
but upward and downward gas motions are driven by the onset of an RHD instability
at later times. 
The middle panel shows the growth of the density perturbations on small length scales.
The density pattern of upflow (''spike") and downflow (''bubble") structure,
which is very reminiscent of RT instability, is clear there.
It is known that the small-scale perturbation has the fastest growth rate for
radiation RT instability, as is the case for pure hydrodynamic RT instability  
(\cite{JacKru11}). We confirm the same tendency in our case, 
in which the gravitational force is of minor importance.
The bottom panel shows the multi-mode phase of the instability.
We found that the mode with a particular wavelength is dominant.
We will show later that this is purely radiation effects.

To see how radiation works to suppress longer-wavelength patterms,
we illustrate the magnified view of density contours around 
one particular optically thick spike in Figure \ref{frad}.
Obviously, the arrows, which indicate the radiation force per unit volume,
in the denser region point toward the low density bubble. 
That is, photons from the dense region diffuses towards the low density region
and produces a correlation between the gradient of the matter density and the radiation flux. 
As a result, at the interface between the two layers
the radiation-pressure force of the dense spike balances
the gas-pressure force of the hot density bubble.
This indicates that radiation processes tend to damp the RHD instability 
on length scales longer than the photon mean free path.
We thus expect that the dense spike has a typical length scale of $\sim \ell$,
which is the seed of the clumps found in Paper~I.
In the optically thin region ($z \gtsim 0$), conversely,
there is no such correlation between the gradient of the matter density 
and the radiation flux, since radiation flux is freely streaming out with no interactions.

\subsection{Correlation Analyses}
In Paper~I
we found remarkable features of the clumpy outflow through the
auto- and cross-correlation analyses of the matter and radiation density distributions.
See if the present clumpy structure shares the same features 
with those found by Paper~I,
we repeat the same analyses but using the present RHD simulation data.
The correlation function 
$C(L)$ is useful 
to find coherent lengths and/or repetition intervals with an interval $L$,
if they exist. 
It is explicitly written as
\begin{eqnarray}
  C(L)  = \left\{ \begin{array}{ll}
    \frac{ \displaystyle 
       \int_{x_1}^{x_2-|L|} \delta f(x+|L|) \delta g(x) dx
    }
                  { 
                  \displaystyle \sqrt{
                    \displaystyle \int_{x_1}^{x_2} {\delta f(x)}^2 dx
                    \displaystyle \int_{x_1}^{x_2} {\delta g(x)}^2 dx
                   }
                   }
     & (L < 0) \\
     \frac{ \displaystyle 
      \int_{x_1}^{x_2-L} \delta f(x) \delta g(x+L) dx
     }
                  { 
                  \displaystyle \sqrt{
                    \displaystyle \int_{x_1}^{x_2} {\delta f(x)}^2 dx
                    \displaystyle \int_{x_1}^{x_2} {\delta g(x)}^2 dx
                   }
                   }
     & (L \ge 0),
  \end{array} \right.
\label{corr}
\end{eqnarray}
where
\begin{eqnarray}
\delta f(x) = f(x) - \average{f},
\end{eqnarray}
\begin{eqnarray}
\delta g(x) = g(x) - \average{g},
\end{eqnarray}
and $\average{f}$ and $\average{g}$ are 
the horizontal average values of $f(x,z)$ and $g(x,z)$, 
respectively.

In order to evaluate the typical size of the clumps made by the instability,
we show auto-correlation functions (ACFs) of the matter density in
the left panel of Figure \ref{acfccf} as a function of typical heights 
of the bottom panel of Figure \ref{evolution}.
The interval $\delta x$ is normalized by the photon mean free path $\ell$.
The width of the central peak represents the typical clump size,
whereas the separation from its neighboring peaks represents the typical clump interval.
The ACFs are integrated by the depicted range of Figure \ref{evolution},
$(x_1, x_2) = (-5, 5)$.
The black line indicates the ACFs
of the RHD instability in the optically thick region ($\rho \sim \rho_0$).
The typical wavelength of the RHD instability corresponds to one optical depth,
\begin{eqnarray}
\tau \sim  \kappa_{\rm es} \rho_0 \ell = 1.
\end{eqnarray}
The blue line indicates the ACF
of the RHD instability in the optically thin region ($\rho \sim 10^{-2}\rho_0$).
This typical wavelength is shorter than those in the optically thick region
(indicated by the black line).
This indicates that the damping force for expanding the dense spikes
is weak in the optically thin region.
Note that the clump size ($\sim \ell$) corresponds to $\sim$ 50 times larger than the grid spacing. 
We can thus conclude that each pattern is well resolved in our simulations.

\begin{figure*}[!t]
\begin{center}
\FigureFile(85mm,55mm){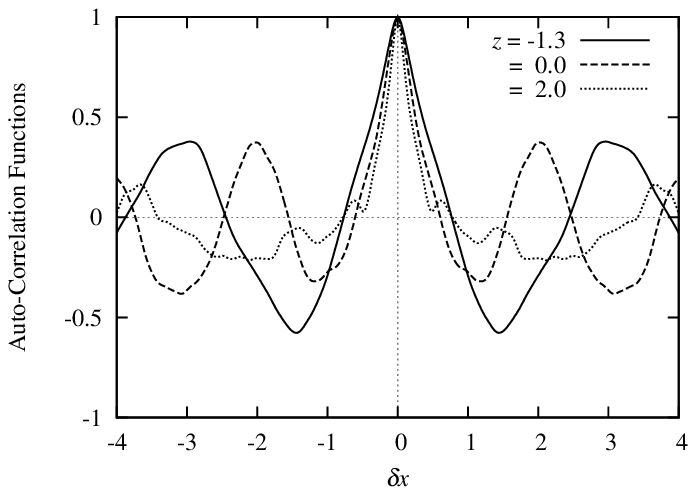}
\FigureFile(85mm,55mm){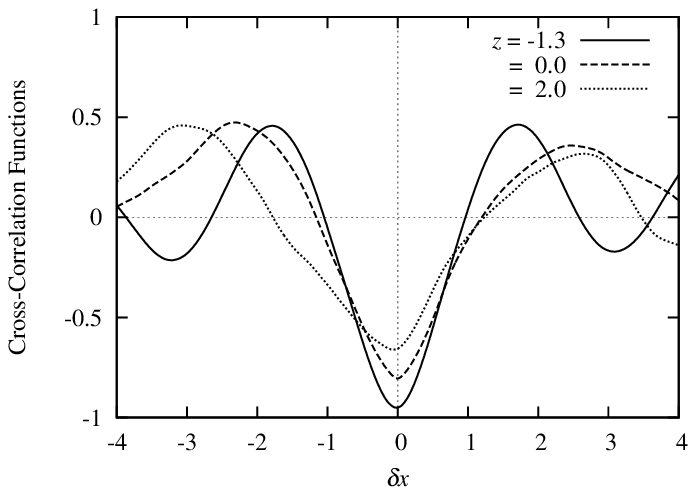}
\caption{
Correlation functions of typical physical values
of the bottom panel of Figure \ref{evolution}.
Left panel:
Auto-correlation functions (ACFs) of the matter density
in the horizontal direction ($\delta x$) as a function of the height ($z$).
The width of the central peak represents the typical wavelength of the RHD instability.
Right panel:
Cross-correlation functions (CCFs) between 
the gas pressure and the radiation pressure.
The negative value at zero separation means the anti-correlation between quantities.
}
\label{acfccf}
\end{center}
\end{figure*}

To see the mechanism of the damping in the RHD instability,
we plot in the right panel of Figure \ref{acfccf} 
the cross-correlation functions (CCFs) 
between the gas pressure and the radiation pressure, $p_{\rm rad}$ ($= E_0/3$),
as a function of the height of the bottom panel of Figure \ref{evolution}.
In the optically thick region (black line),
the anti-correlation between the radiation pressure and the gas pressure is clear.
The matter density 
is correlated with the radiation pressure
and anti-correlated with the gas pressure.
In the optically thin region (blue line), on the other hand,
the profile of the CCF becomes diffuse,
since the radiation flow has a free streaming 
which is independent on the matter density pattern.
Thus, the optical thick, radiating fluids form the typical structure with $\tau \sim 1$
by the RHD instability,
while the optically thin fluids has no typical structure. 
We note that the ram pressure is smaller than the gas pressure.

\section{Discussion}
\subsection{The Nature of the Radiation Hydrodynamic Instability}
We simulated evolution of the RHD instability 
in plane-parallel, super-Eddington atmosphere,
finding that an instability grows and forms a characteristic spatial pattern
whose wavelength corresponds to one optical depth.
Although the simulated instability is not purely of a sort of the RT instability,
since gravity is not essential in the present case,
it shares some similarities with the radiation RT instability.
A big distinction exists, however; that is the size of clumps made by the instability.
The previous analytic and numerical studies of radiation RT instability
show no such typical scales for unstable modes
for the linear growth rate
(\cite{JacKru11}; \cite{Jia+13}).

In the optically thin limit,
radiation acts as a part of an effective gravitational field;
$\mbox{\bf g}_{\rm eff} \equiv \mbox{\bf g} + \chi \bm{F_0}/c 
= -\mathrm{g}_{\rm eff} \bm{e}_z$.
In the optically thick limit,
the fluid is expressed as one fluid because of strong coupling between matter and radiation,
and the dispersion relation is reduced to that of the pure hydrodynamic RT instability.
The dispersion relation of both limit is approximately expressed by
\begin{eqnarray}
  \omega \propto \left\{ \begin{array}{ll}
    \displaystyle  k^{1/2}
     & \displaystyle ( \lambda  \ltsim H) \\
     \displaystyle  k
     & \displaystyle ( \lambda  \gtsim H ).
  \end{array} \right.
\label{corr}
\end{eqnarray}
where $\omega$ is the growth rate,
$k$ is the wave number,
$\lambda$ is the wavelength,
and $H$ is the pressure scale-height.
The steep power of the dispersion relation in the long-wavelength ($\lambda  \gtsim H$)
is explained by the compression effects.

Then, why can the RHD instability which we encounter here
exhibit a typical length scale?
The reason is found in the anti-correlation diagram between the gas pressure 
and the radiation pressure.
Let us consider the dispersion relation of the RHD equation.
By using the linearized values of $A(z) \exp{ i(kx-\omega t)}$,
the radial component of the equation (\ref{mom}) is derived as follows:
\begin{eqnarray}
-i \omega \rho_0 \delta v_x = -ik\delta p_{\rm gas} 
  + \kappa_{\rm es} \rho_0 \delta F_0^x.
\label{linear}
\end{eqnarray}
Here, we note the $k$-dependence of the optical depth, $\tau \propto k^{-1}$.
Also note that one pressure scale-height roughly corresponds to one optical depth
just below the photosphere,
\begin{eqnarray}
H \sim \ell.
\label{photo}
\end{eqnarray}
That is to say, 
the layer in question is optically thin in the short wavelength limit ($\lambda \to 0$),
while it is optically thick in the long wavelength limit ($\lambda \to \infty$).
We thus have
\begin{eqnarray}
-i \omega \rho_0 \delta v_x  =
   \left\{ \begin{array}{ll}
    -ik\delta p_{\rm gas}
     & \displaystyle ( \lambda  \ltsim  \ell) \\
     -ik (\delta p_{\rm gas} + \delta p_{\rm rad}) 
     & \displaystyle ( \lambda  \gtsim  \ell),
  \end{array} \right.
\label{rrti}
\end{eqnarray}
since the horizontal radiation force is zero in the optically thin case.
From the right panel of Figure \ref{acfccf}, we found the following relation:
\begin{eqnarray}
\delta p_{\rm gas} \sim - \delta p_{\rm rad}.
\label{cond}
\end{eqnarray}
Because of this relation the right-hand-side (R.H.S.)
in equation (\ref{rrti}) vanishes in the long wavelength limit.
We finally have the following form of the dispersion relation
for the RHD instability in super-Eddington atmosphere,
 \begin{eqnarray}
  \omega \propto \left\{ \begin{array}{ll}
    \displaystyle  k^{1/2}
     & \displaystyle ( \lambda  \ltsim \ell) \\
     \displaystyle  0
     & \displaystyle ( \lambda  \gtsim \ell).
  \end{array} \right.
\label{disp}
\end{eqnarray}
We now understand that
why the growth of perturbations on longer scales than $\ell\sim H$ is damped.
That is due to radiation damping.

It is essential that 
this instability exhibits a typical length scale of $\lambda \sim \ell$
only when it occurs near the photosphere in the marginally optically thick system. 
Longer wavelength perturbations suffer from the damping by radiation pressure,
since then the growth timescale becomes longer than 
the sound crossing time so that the decoupling between matter and radiation occurs. 
In other words,
this instability does not occur in deep layers of super-Eddington objects,
such as supercritical accretion flow and massive star.
Within optically thin layers, in contrast, an instability can still occur
but with no characteristic length scales. 
This is because radiation damping mechanism does not occur under the optically thin condition.

We have repeatedly stressed that the
RHD instability in a super-Eddington atmosphere 
has the typical wavelength of $\tau \sim 1$ 
based on the simulations made with the FLD approximation.
but we should point that the precise size of the clumps may depend 
on the method of transfer calculations.
Even when more sophisticated methods for radiative transfer problems are adopted, 
our main conclusion
will not be altered, since the typical wavelength is created 
in marginally optically thick layers below the photosphere.

\subsection{Comparison with Clumpy Outflow}
In Paper~I 
we performed two-dimensional global
radiation-MHD simulations of supercritical accretion flows
onto black holes,
finding that the outflows have a clumpy structure above the photosphere
and the typical clump size corresponds to about one optical depth.
Radiation RT instability was suspected to be the most plausible
cause of clump formation, 
since the clumpy structure appears 
in the layer where the upward radiation force is superior to the downward gravity force.
Let us check the condition of the RHD instability of equation (\ref{cond}).
We checked the CCFs between 
the radiation pressure and the gas pressure in the clumpy outflow region in Paper~I, 
confirming a clear tendency of anti-correlation between them.
Thus the clumps keep their shapes by the radiation pressure against the gas pressure
by the RHD instability.
As found in the previous simulations,
each clump dynamically changes its shape in time,
which would be the nonlinear effects.

\subsection{Other Radiation-Hydrodynamic Instabilities}
There are other RHD instabilities creating an inhomogeneous (porous) structure
of the matter density in radiation-pressure supported atmosphere
(\cite{PreSpi73}; \cite{SpiTao99}; \cite{Sha01}; \cite{BlaSoc03}; \cite{Fuk03}; \cite{Tur+05}).
\citet{Sha01} made a global linear stability analysis for the optically thick, 
radiation-dominated atmospheres in massive stars,
finding the unstable mode close to the Eddington limit
(see also \cite{Sha00}).
As \citet{Jia+13} mentioned, however, 
the instability criteria of the RHD instabilities are not completely understood yet. 
In fact, our RHD nor radiation-MHD simulations did not find an
anti-correlation, as was claimed by \citet{Sha01},
but a correlation between the matter and radiation-energy densities.

The reason for this discrepancy may stem from the fact that
\citet{Sha01} examined a stability of media with no initial radiation fields,
while we started simulations with non-zero radiation energy density.
As we have seen in Figure \ref{frad}
radiation flux (and radiation force asserted) from optically thick regions to thin regions
is essential to the RHD instability of a sort that we are encountered with in the present study.
We also confirm exactly the same tendency in the simulation data of Paper~I.
The difference may also arise because of different physical conditions
postulated in these studies;
i.e., this work, \citet{JacKru11},
\citet{KruTho12}, and \citet{Jia+13} focus on interface instabilities,
such the Rayleigh-Taylor instability and its variations,
while \citet{Sha01} considered a global instability
and \citet{BlaSoc03} examined local instabilities.

\subsection{Observational Implications}
The present analysis could be effective not only to luminous black hole objects
but also to the atmosphere of massive stars.
The most notable consequence of this instability is time variability due to absorption by optically thick clumps (Paper~I). 
Shaviv (2001) discussed applications to various super-Eddington sources,
such as luminous blue variables (LBVs), Wolf-Rayet stars, classical novae, 
and so on (see also \cite{OwoSha12}). 
He conjectured that clumpy outflow should be prominent
features of super-Eddington sources, and this is what we concluded through RHD simulations.

\section{Conclusion}
In this paper, we examined 
the property of the RHD instability in a 
plane-parallel, super-Eddington
atmosphere 
by using two-dimensional RHD simulations to understand the mechanim
underlying clump formation in super-Eddington outflow found in Paper~I.
Here are our new findings:
\begin{itemize}
\item
An RHD instability in super-Eddington atmosphere
shows a typical length scale for clumps, which corresponds to one optical depth,
after the growth of perturbations on small scale.
\item 
There is a clear correlation between the matter density and the radiation pressure,
and an anti-correlation between the matter density and the gas pressure 
in the optically thick region.
This radiation damping is responsible for suppression of the growth of longer wavelength
perturbations than the typical clump size.
\item 
The clumps found in Paper~I have the same feature of
the anti-correlation between the radiation pressure and the gas pressure.
\item
We find no evidence of the occurrence of instability 
of a sort that \citet{Sha01} reported, which shows an 
anti-correlation between the matter density and the radiation energy density.
The reason seen to reside in the difference of the initial condition of radiation field.
\end{itemize}

\bigskip
We would like to thank M. Umemura for useful comments and discussions.
We also thank the anonymous referee for comments to improve the paper.
This work is supported in part by Grants-in-Aid of the Ministry of Education, Culture,
Sports, Science and Technology (MEXT) (24740127, KO) 
and by a Grant-in-Aid for the Global COE
Programs on gThe Next Generation of Physics, Spun from
Diversity and Emergenceh from MEXT (SM). 
Numerical computations were in part carried out on Cray XC30 at
Center for Computational Astrophysics (CfCA) of National
Astronomical Observatory of Japan.

\end{document}